# The unresolved mystery of the great divergence is solved

Ron W Nielsen[1]

Environmental Futures Research Institute, Gold Coast Campus, Griffith University, Qld, 4222, Australia

**Abstract.** The so-called great divergence in the income per capita is described in the Unified Growth Theory as the mind-boggling and unresolved mystery about the growth process. This mystery has now been solved: the great divergence never happened. It was created by the manipulation of data. *Economic growth in various regions is at different levels of development but it follows similar, non-divergent trajectories.* Unified Growth Theory is shown yet again to be incorrect and scientifically unacceptable. It promotes incorrect and even potentially dangerous concepts. The distorted presentation of data supporting the concept of the great divergence shows that economic growth is now developing along moderately-increasing trajectories but mathematical analysis of the same data and even their undistorted presentation shows that these trajectories are now increasing approximately vertically with time. So, while the distorted presentation of data used in the Unified Growth Theory suggests generally sustainable and secure economic growth, the undistorted presentation of data demonstrates that the growth is unsustainable and insecure. The concept of takeoffs from stagnation to the sustained-growth regime promoted in the Unified Growth Theory is also dangerously misleading because it suggests a sustainable and prosperous future while the mathematical analysis of data shows that the current economic growth is insecure and unsustainable.

## Introduction

Those who are less familiar with the scientific process of investigation might not be aware that there is also unscientific approach, which unfortunately appears to be used sometimes even in academic circles. It is important to have a clear understanding of these two different ways of investigation in order to be able to distinguish between acceptable and unacceptable claims and conclusions.

In science, theories are tested by data. In unscientific discussions, data are tested by theories. In unscientific presentations, selective use of data is common. Data are manipulated, distorted or rejected if they do not agree with preconceived ideas.

In the scientific research, contradicting evidence is not only accepted but looked for because it usually leads to new discoveries. In unscientific discussions, contradicting evidence is studiously rejected because it threatens the established knowledge.

In science, data are rigorously analysed. In non-scientific discussions, rigorous analysis is avoided and interpretations of data are based on impressions, but

---

impressions can be misleading. The Sun appears to be circling around the Earth but it is not. Logical explanations are not necessarily reliable.

It is also important to understand the limitations of mathematics. Elaborate stories and explanations can be translated into mathematical language but such translations are meaningless unless they can be tested by data.

We should never be mesmerised by complicated mathematical formulae and presentations. The essential question is whether the presented mathematics can be tested by relevant data. If stories translated into mathematics cannot be tested (or are not tested) by data, if they have to be accepted by faith, then obviously they have no scientific value and they can be (or even should be) rejected. Mathematical formulations should be making testable predictions. A story dressed up in a mathematical gown will be just a story unless it makes a testable prediction.

A good example of the unscientific approach to research is the Unified Growth Theory (Galor, 2005a, 2011). Data are manipulated and distorted. Selected data, which appear to support preconceived concepts, are repeatedly quoted. Excellent data of Maddison (2001) are used during the formulation of this theory but they are never analysed. They are presented in distorted ways to support the preconceived ideas. Galor translates his assumed and scientifically-unsupported interpretations of economic growth into many complicated but rather primitive mathematical formulae. However, he does not make even a single mathematical prediction, which can be tested directly by data. His mathematical expressions do not describe growth trajectories that could be compared with data he uses during the formulation of his theory.

His concepts can be only tested indirectly by showing that within the range of the mathematically-analysable data there was no stagnation, no sudden takeoffs, no "remarkable" or "stunning" escapes from the Malthusian trap (Galor, 2005a, pp. 177, 220) and no transition from stagnation to the so called sustained growth regime (Nielsen, 2014, 2015a, 2015b, 2016a, 2016b, 2016d, 2016e, 2016f, 2016g, 2015h). Economic growth in the past was sustainable and secure, as indicated by the steadily-increasing hyperbolic trajectories, but now it is unsustainable and insecure (Nielsen, 2015b). The numerous mathematical formulae used in the Unified Growth Theory do not describe or explain the historical economic growth because they incorporate concepts, which are either contradicted repeatedly by data or have to be accepted by faith.

We have already demonstrated (Nielsen, 2014, 2015a, 2016a, 2016b, 2016d, 2016e, 2016f, 2016g, 2016h) that Galor's Unified Growth Theory is fundamentally incorrect because it is based on fundamentally-incorrect ideas. We have shown that within the range of the mathematically-analysable data, historical economic growth and the growth of population were hyperbolic. For the economic growth, the range of evidence is limited but for the growth of human population it can be extended to 10,000 BC (Nielsen, 2016b). We have demonstrated that within the range of analysable data, there was no Malthusian stagnation and no Malthusian trap. The growth was slow over a long time but it was steadily increasing and there was no transition at any time in the past that could be described as a sudden takeoff, spurt, sprint or explosion. We have demonstrated that Galor's claim of sudden takeoffs is repeatedly contradicted by data. There were no takeoffs and consequently there was also no differential timing of takeoffs. During the time of claimed takeoffs, economic growth and the growth of population were continuing to increase along undisturbed and remarkably stable hyperbolic trajectories until recently, when they started to be diverted to slower but still fast-increasing trajectories. This conclusion applies not only to the growth of the Gross Domestic Product (GDP) and population but also to the growth of income per capita (GDP/cap). We do not have to try to explain the mechanism of the epoch of Malthusian stagnation and of the escape from the Malthusian trap because there was no stagnation and no trap. What we have to explain is why the growth in the past was hyperbolic, why it was so remarkably stable and why it started to be diverted recently to new, non-hyperbolic trajectories.

## The concept of the great divergence

The concept of the great divergence belongs to a set of other phantom "mysteries about the growth process" (Galor, 2005a, p. 220) invented by Galor and reinforced by the habitually distorted presentations of data (Ashraf, 2009; Galor, 2005a, 2005b, 2007, 2008a, 2008b, 2008c, 2010, 2011, 2012a, 2012b, 2012c; Galor and Moav, 2002; Snowdon & Galor, 2008). One example of the distorted presentation of data used routinely by Galor is shown in Figure 1. In contrast, the accurate presentation of precisely the same data, together with their mathematical analysis, is shown in Figure 2.

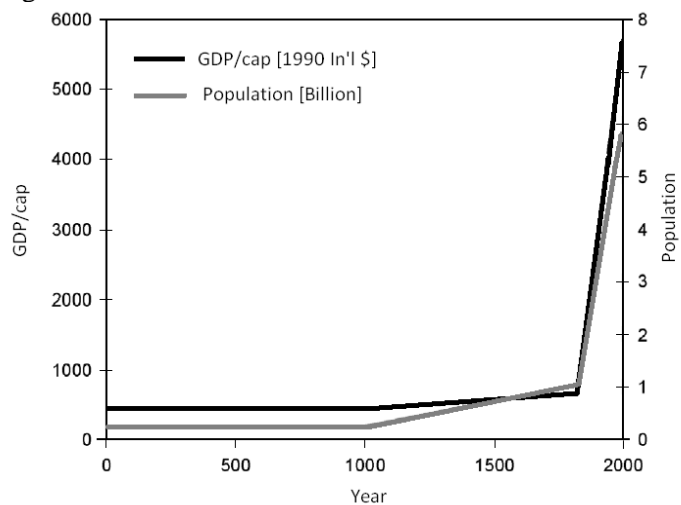

**Figure 1.** *Example of the ubiquitous, grossly-distorted and self-misleading diagrams used to create the Unified Growth Theory (Galor, 2005a, 2011). Maddison's data (Maddison, 2001) were used during the formulation of this theory but they were never analysed. Such state-of-the-art was used to construct a system of scientifically-unsupported interpretations, explanations and "mysteries of the growth process" (Galor, 2005a, p. 220).*

In the distorted and appropriately manipulated presentation of data shown in Figure 1 we can see clearly the non-monotonic growth of population and of the GDP/cap. After the apparent long stagnation, we see a sudden takeoff to a new regime of growth. Galor made no attempt to analyse data, which is surprising because their analysis is trivially simple (Nielsen, 2014). The manipulated data appear to support the concept of stagnation and takeoffs described usually as the escape from the Malthusian trap.

In contrast, the accurate display of precisely the same data suggests entirely different interpretation. General features presented in Figure 1 are still maintained but now mathematical analysis of these data shows that the GDP/cap and the size of the population were increasing monotonically (Nielsen, 2014, 2015a, 2016a, 2016d, 2016g). There were no sudden takeoffs from stagnation to growth because there was no stagnation and because the acceleration was gradual along the entire range of these distributions. The gradient and the growth rate of the GDP/cap distribution were changing monotonically without any discontinuity, which could be claimed as a takeoff (Nielsen, 2015a).

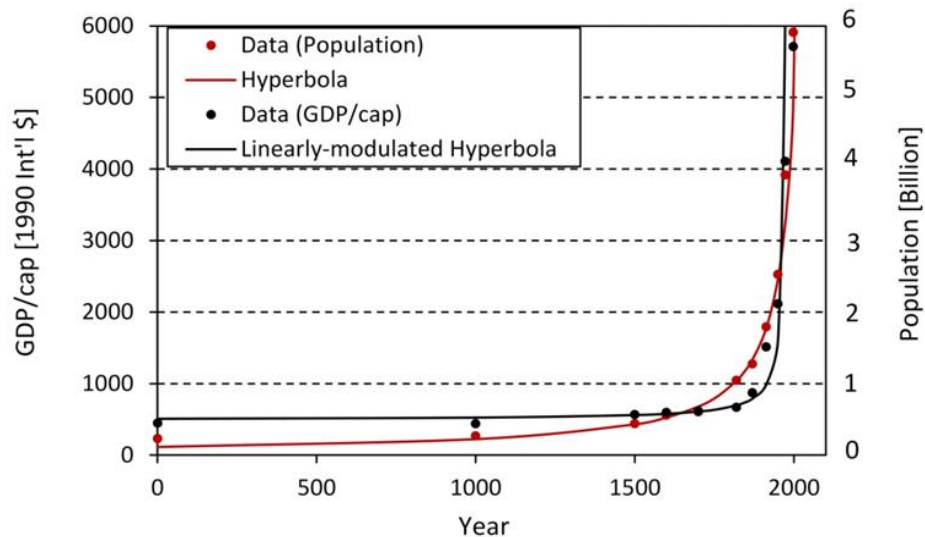

**Figure 2**. *The same data (Maddison, 2001) as used in Figure 1but now displayed accurately and analysed. They follow monotonically-increasing distributions, which cannot be divided into distinctively-different components (Nielsen, 2014, 2015a, 2016a, 2016d, 2016g).*

Even though the GDP/cap distribution seems to suggest a sudden increase, this feature is just an illusion, which is dispelled by the mathematical analysis of data or even by simply using semilogarithmic scales of reference (Nielsen, 2015a, 2016g). The GDP/cap is the ratio of two distributions, the distribution describing the growth of the GDP and the distribution describing the growth of population. Both were increasing hyperbolically and monotonically (Nielsen, 2015a, 2016a, 2016d). The displayed features (slow growth over a long time and fast growth over a short time) represent nothing more than mathematical properties of monotonically-increasing hyperbolic distributions. They are not the unique properties of economic growth but economic growth happens to be hyperbolic.

It is impossible to locate a transition from the slow to fast growth for hyperbolic distributions (Nielsen, 2014) because such a transition does not exist. The GDP/cap distributions are simply the linearly-modulated and monotonically-increasing hyperbolic distributions (Nielsen, 2015a).

The distorted diagram used by Galor to support his erroneous concept of the great divergence is presented in Figure 3. This distorted presentation of Maddison's data was reproduced from Galor's publication (Galor, 2005a, p. 175). It shows that over a long time there was hardly any difference in the economic growth for various regions. However, from around the time of the Industrial Revolution, 1760-1840 (Floud & McCloskey, 1994), there was a sudden takeoff and the economic growth diverged into distinctly different trajectories.

We have already demonstrated that there were no takeoffs in the growth of the GDP and GDP/cap (Nielsen, 2015a, 2016e, 2016g) and consequently there was also no differential timing of takeoffs claimed by Galor in his Unified Growth

Theory (Galor, 2005a, 2011). We have also demonstrated that there were no takeoffs in the growth of human population in the past 12,000 years (Nielsen, 2016b, 2016d). The incorrectly-claimed takeoffs represent just the natural continuations of hyperbolic growth. Analysis of data shows that at the time of the alleged takeoffs and in clear contradiction of the Unified Growth Theory, economic growth in various regions was either continuing to increase along undisturbed hyperbolic trajectories or started to be diverted to *slower* trajectories.

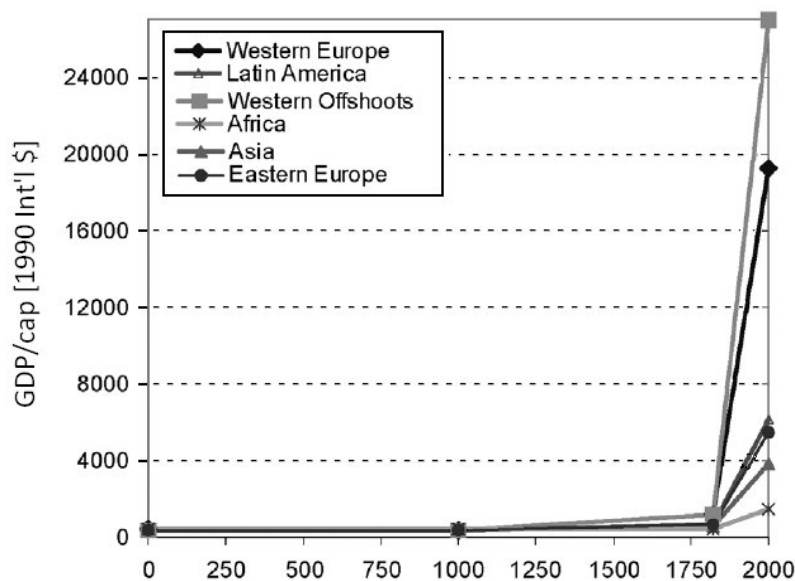

**Figure 3**. *A typical distorted presentation of Maddison's data (Maddison, 2001) used by Galor to support his concepts of takeoffs and of the great divergence (Galor, 2005a, p. 175).*

Now we shall show that there was no great divergence in income per capita. We shall show again that the Unified Growth Theory is scientifically unacceptable. It does not describe the mechanism of economic growth. It describes phantom features constructed by the manipulation of data.

We shall show that the great divergence never happened. However, we shall also *explain* how Galor constructed his great divergence. We shall show how the great divergence can be constructed by a distorted presentation of any distributions, which increase slowly over a long time and fast over a short time. They do not have to be distributions describing economic growth.

## The early data of Maddison

We shall first investigate precisely the same data (Maddison, 2001) as used by Galor (2005a, 2011) during the formulation of his Unified Growth Theory and we shall show that they do not support the concept of the great divergence. Results of mathematical analysis of these data are shown in Figures 4-9. Parameters of the fitted distributions have been listed earlier (Nielsen, 2016g). The fitted curves are the linearly-modulated hyperbolic distributions (Nielsen, 2015a) obtained by dividing hyperbolic distributions fitting the corresponding GDP and population data (Nielsen, 2016a, 2016d). All GDP/cap values are in 1990 International Geary-Khamis dollars.

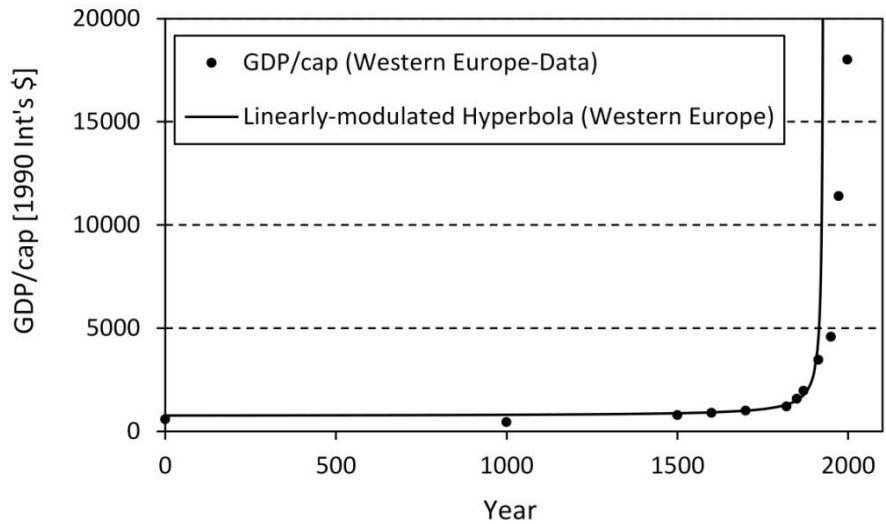

**Figure 4**. *Growth of income per capita, i.e. Gross Domestic Product per capita (GDP/cap), in Western Europe (Maddison, 2001; Nielsen, 2016g). From around 1913, economic growth in Western Europe started to depart from the historical linearly-modulated hyperbolic distribution. However, it continued to increase close to the historically-predicted trajectory.*

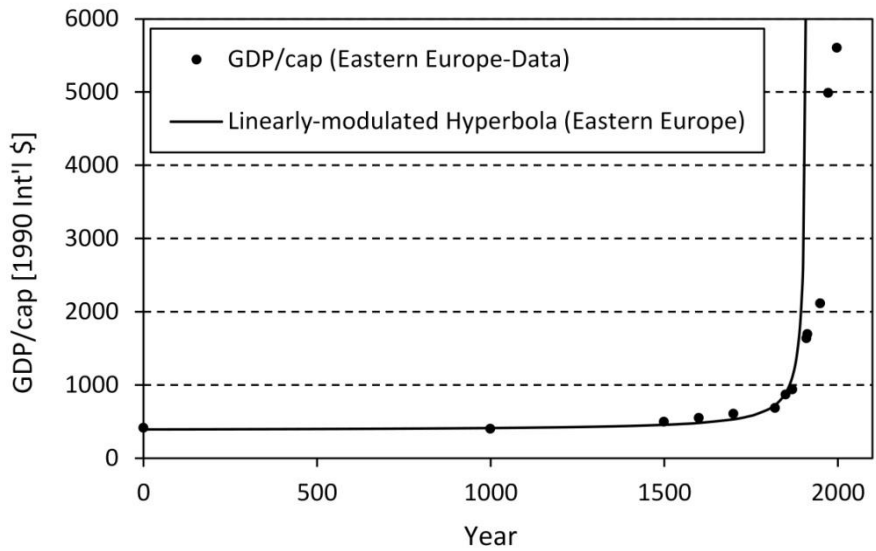

**Figure 5**. *Growth of income per capita in Eastern Europe (Maddison, 2001; Nielsen, 2016g). From around 1870, economic growth in Eastern Europe started to depart from the historical linearly-modulated hyperbolic distribution. However, it continued to increase close to the historically-predicted trajectory. The growth was not diverted to a distinctly different and gently-increasing trajectory as claimed by Galor (2005a, 2011; cf Figure 3).*

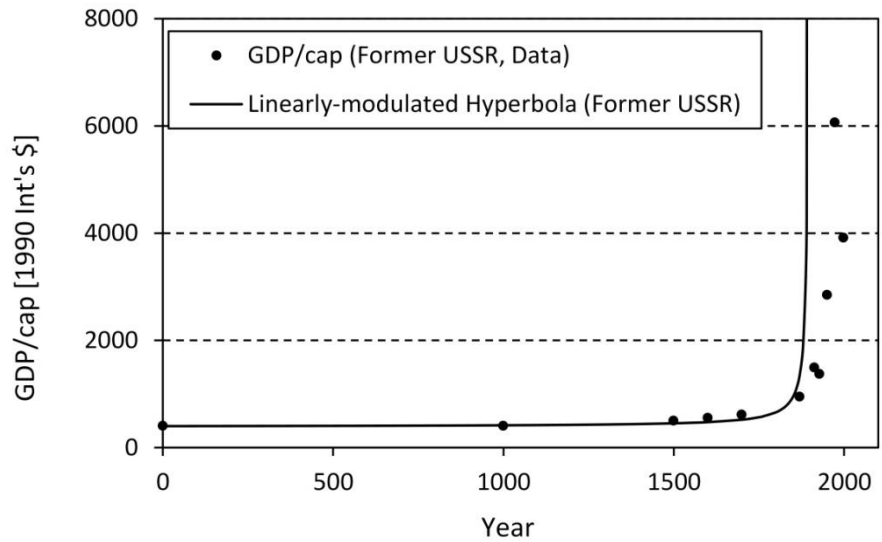

**Figure 6**. *Growth of income per capita in countries of the former USSR (Maddison, 2001; Nielsen, 2016g). From around 1870, economic growth in the former USSR started to depart from the historical linearly-modulated hyperbolic distribution. However, it continued to increase close to the historically-predicted trajectory.*

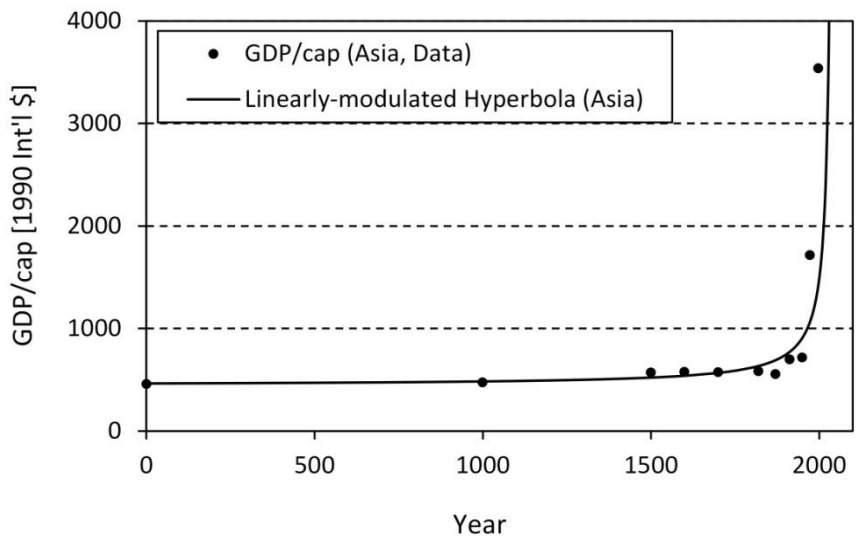

**Figure 7**. *Growth of income per capita in Asia (Maddison, 2001; Nielsen, 2016g). The data follow closely the linearly-modulated hyperbolic distribution. There was no divergence to a distinctly different and gently-increasing trajectory as claimed by Galor (2005a, 2011; cf Figure 3).*

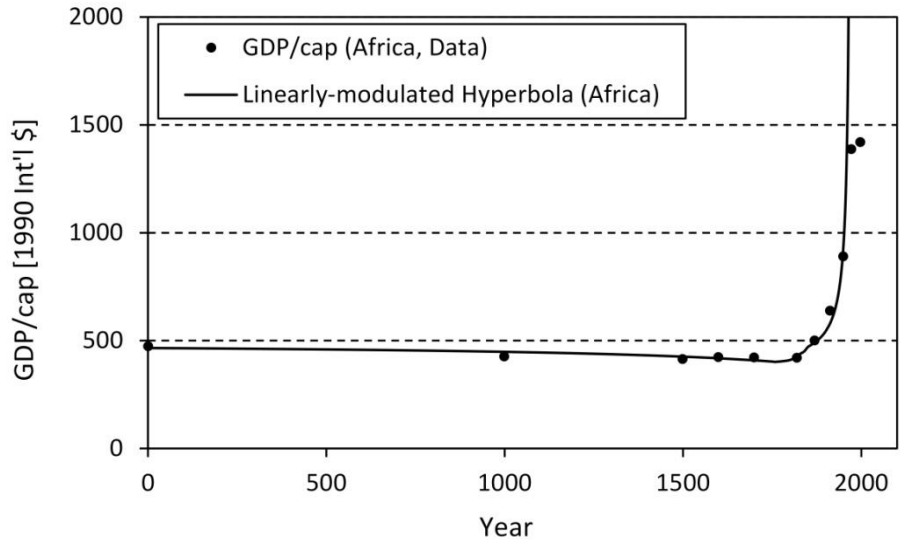

**Figure 8**. *Growth of income per capita in Africa (Maddison, 2001; Nielsen, 2016g). The data follow closely the linearly-modulated hyperbolic distributions. There was no divergence to a distinctly different and gently-increasing trajectory as claimed by Galor (2005a, 2011; cf Figure 3).*

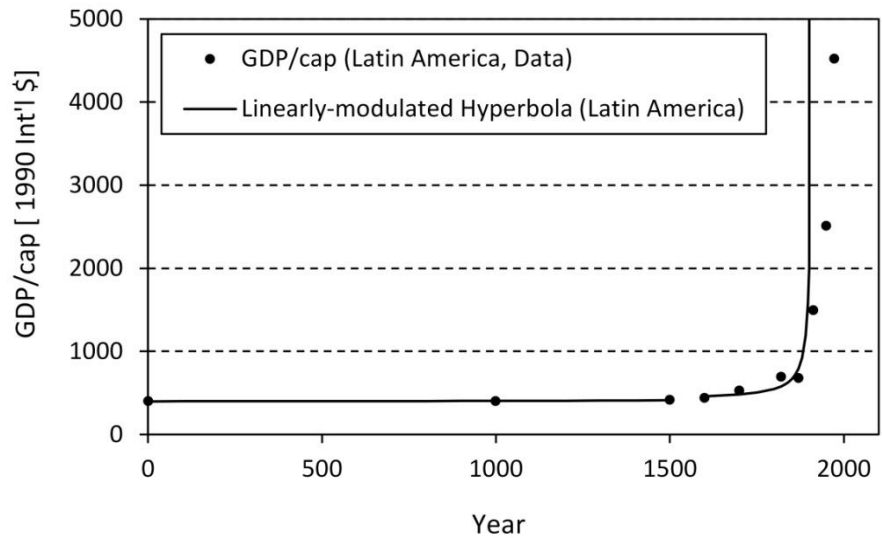

**Figure 9**. *Growth of income per capita in Latin America (Maddison, 2001; Nielsen, 2016g). From around 1913, economic growth in Latin America started to depart from the historical linearly-modulated hyperbolic distribution. However, it continued to increase close to the historically-predicted trajectory. The growth was not diverted to a distinctly different and gently-increasing trajectory as claimed by Galor (2005a, 2011, cf Figure 3).*

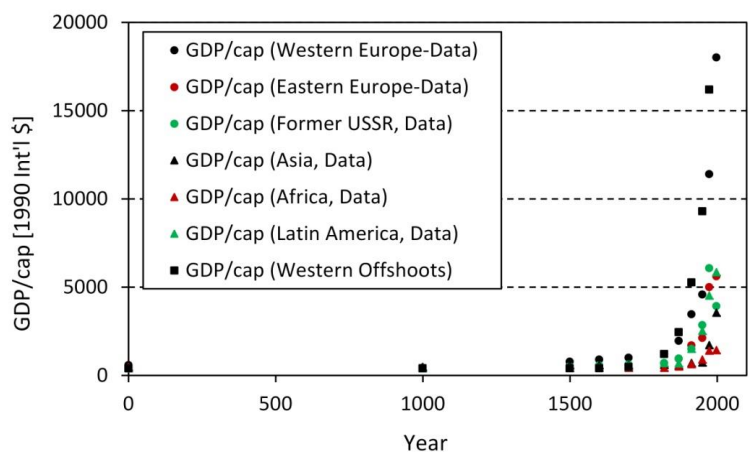

**Figure 10**. *Growth of income per capita in all regions, including Western offshoots (Maddison, 2001). They are all increasing in approximately the same direction. There is no divergence to distinctly different trajectories.*

The data for Western Offshoots were not analysed because of their poor quality, but they are displayed in Figure 10. Their economic growth is similar to the growth in Western Europe in the sense that they are clearly ahead of other regions. However, distributions presented in Figures 4-9 show that economic growth in all regions follows similar trajectories. The difference between regions is not in their divergence to distinctly different trajectories as claimed incorrectly by Galor but in their *levels* of economic development.

Distributions presented in Figures 4-9 are clearly different than the distorted distributions constructed by Galor and presented in Figure 3. In Galor's distorted presentation of data there is a cluster of regions (Eastern Europe, Asia, Africa and Latin America) whose economic growth follows distinctly different trajectories than the growth in Western Europe. This information is incorrect because the analysis of precisely the same data shows clearly that all distributions are similar, including the distribution representing the economic growth in Africa. They are all following similar trajectories with a common tendency to increase nearly vertically and close to the historical linearly-modulated hyperbolic trajectories.

The common characteristic feature of all these empirical distributions shown in Figures 4-9 is that they have changed gradually from being nearly horizontal to nearly vertical. We shall show later that when such distributions become nearly vertical it is easy to distort them and construct the great divergence, and it does not matter whether they follow the fitted linearly-modulated hyperbolic distributions or not.

The contrast between Maddison's data and their distorted image constructed by Galor is particularly clear if we compare Figure 3 with Figure 8. In Figure 3, the data for Africa follow a gently-increasing trajectory after around 1800, i.e. a trajectory characterised by a small gradient. The correct display of the same data presented in Figure 8 shows diametrically opposite features: the data for Africa follow a steep trajectory, i.e. the trajectory characterised by a large gradient. This trajectory is approximately vertical.

In Galor's distorted presentation of data the trajectory for Africa after around 1800 is distinctly different than the trajectory for Western Europe. However, precisely the same data displayed in Figures 4 and 8 demonstrate that the trajectories for Africa and Western Europe are similar. The only difference is that Africa is further behind in its level of development.

In Galor's distorted presentation of data, economic growth in Eastern Europe, Asia and Latin America follows also gently increasing trajectories after around 1800, similar to the trajectory for Africa. However, precisely the same data displayed properly in Figures 5, 7, 8 and 9 show that they all follow approximately vertical trajectories in much the same way as the data for Western Europe. The only difference is again that Western Europe is further ahead but it is further ahead on the virtually the same trajectory.

With such distorted presentation of data it is not surprising that Galor discovered so many "mind-boggling" and "perplexing" "mysteries of the growth process" (Galor, 2005a, pp. 177, 220), mysteries representing phantom features created by the manipulation of data.

In contrast with his distorted presentation of data, the gradient of all empirical trajectories in this section of time is large. They all increase nearly vertically. We should remember that these trajectories do not describe the economic growth represented by the GDP but the GDP *per capita*. Such a growth cannot be explained by the growth of human population. It reflects our surprisingly fast-increasing demands, which propel economic growth along unsustainable trajectories.

Galor's theory conveys dangerously incorrect information. According to his distorted presentation of data shown in Figure 3, income per capita in certain regions (Eastern Europe, Asia, Africa and Latin America) is following gently increasing trajectories after around 1800. Such trajectories are relatively safe. However, the correct presentation of precisely the same data shows that in *all* regions income per capita is following fast-increasing trajectories. The data show that there is now a critical urgency to regulate economic growth but Galor's theory suggests that there is no danger.

## The latest data of Maddison

Data published by Maddison in 2010 show even more clearly that there was no divergence in the economic growth. These data were available to Galor before the publications of his book (Galor, 2011) but unfortunately they were not analysed. Had Galor analysed these data he would have soon discovered many interesting features characterising economic growth, features, which are repeatedly in contradiction with his Unified Growth Theory (Galor, 2005a, 2011).

Results of analysis of these new data (Maddison, 2010) are shown in Figures 11-16. Their combined display is presented in Figure 17.

Galor's "mind-boggling" and "perplexing phenomenon of the Great Divergence" (Galor, 2005a, pp. 177, 220) has now been solved – there is no mystery. This mystery and all other of his mysteries were created by the manipulation of data. In Galor's publications (Galor, 2005b, 2007, 2008a, 2008b, 2008c, 2010, 2012a, 2012b, 2012c; Galor and Moav, 2002; Snowdon & Galor, 2008) data are repeatedly manipulated and presented using distorted diagrams.

The common characteristic feature of Maddison's data describing the growth of income per capita (Maddison, 2001, 2010) in various regions is again that their nearly horizontal trajectories changed gradually into nearly vertical trajectories. They have never diverged into distinctly different trajectories as claimed by Galor (see Figure 3). Even when the economic growth started to be diverted to slower trajectories it continued to increase close to the historical, fast-increasing, linearly-modulated hyperbolic trajectories.

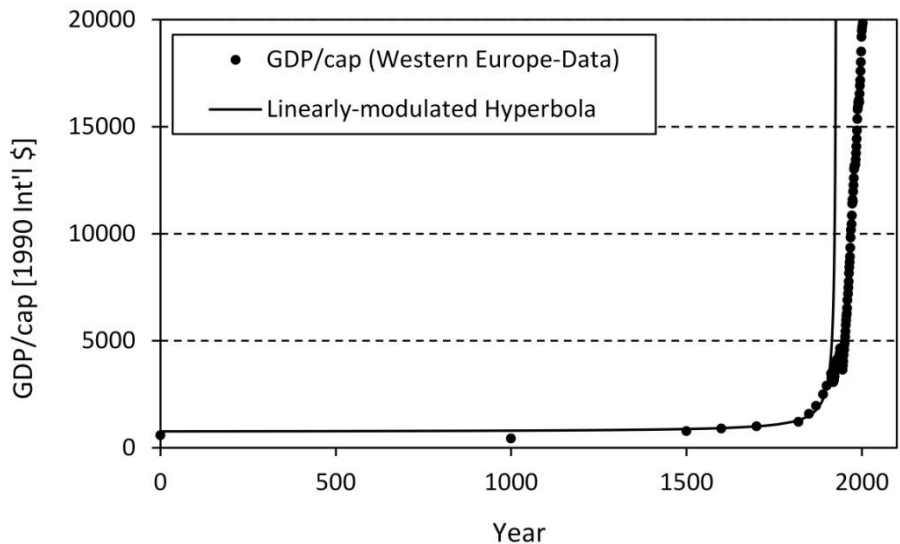

**Figure 11**. *Growth of income per capita in Western Europe (Maddison, 2010; Nielsen, 2016g). Between 1900 and 1913, economic growth in Western Europe started to depart from the historical linearly-modulated hyperbolic distribution. However, it continues to increase close to the historically-predicted trajectory.*

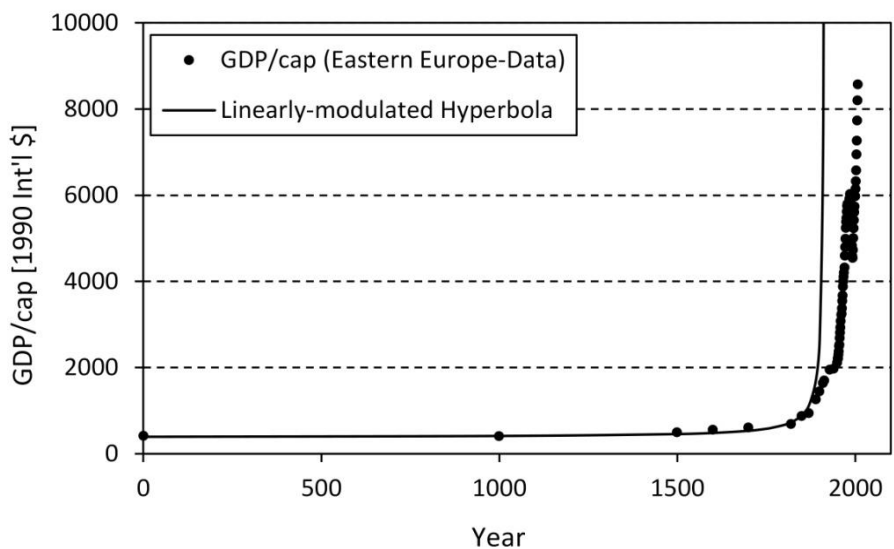

**Figure 12**. *Growth of income per capita in Eastern Europe (Maddison, 2010; Nielsen, 2016g). From around 1850 economic growth in Eastern Europe started to depart from the historical linearly-modulated hyperbolic distribution. However, it continues to increase close to the historically-predicted trajectory. The growth was not diverted to a distinctly different and gently-increasing trajectory as claimed by Galor (2005a, 2011; cf Figure 3).*

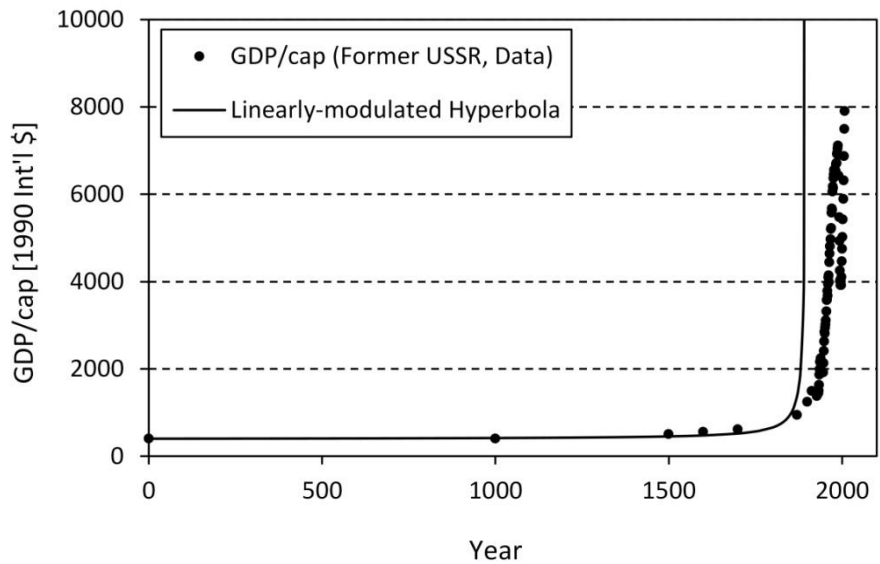

**Figure 13**. *Growth of income per capita in countries of the former USSR (Maddison, 2010; Nielsen, 2016g). Close to around 1870 economic growth in countries of the former USSR started to depart from the historical linearly-modulated hyperbolic distribution. However, it continues to increase close to the historically-predicted trajectory.*

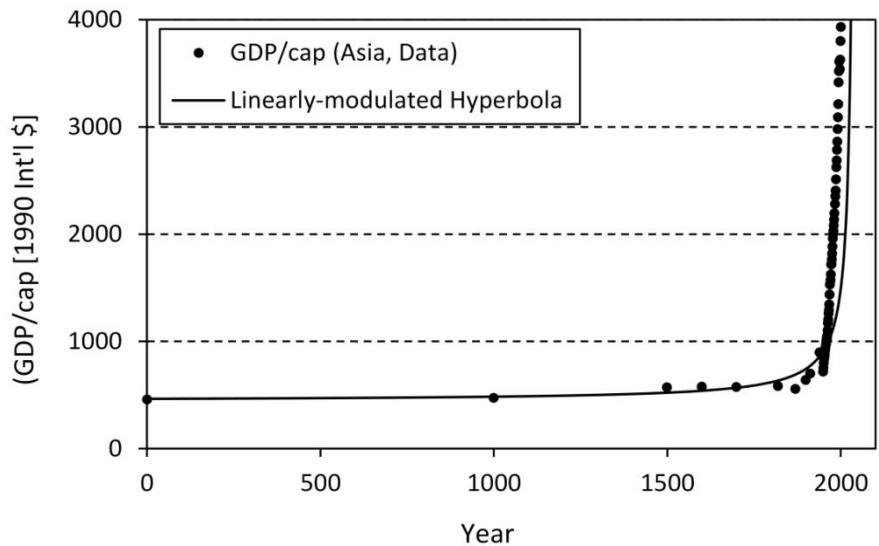

**Figure 14**. *Growth of income per capita in Asia (Maddison, 2010; Nielsen, 2016g). After a brief decline between 1940 and 1950, the growth of income per capita in Asia was diverted to a slightly faster trajectory. However, it continues to increase close to the historically-predicted linearly-modulated hyperbolic distribution. The growth was not diverted to a distinctly different and gently-increasing trajectory as claimed by Galor (2005a, 2011; cf Figure 3).*

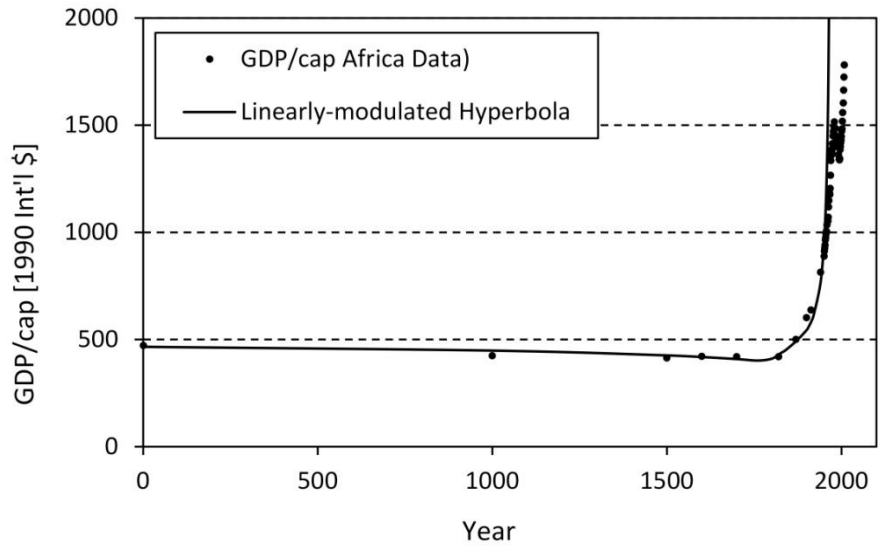

**Figure 15**. *Growth of income per capita in Africa (Maddison, 2010; Nielsen, 2016g). In clear contradiction of Galor's claim supported by his distorted presentation of Maddison's data, the growth of income per capita did not diverge to a slowly-increasing trajectory but continued to increase along a nearly vertical trend close to the historical linearly-modulated hyperbolic distribution (cf Figure 3).*

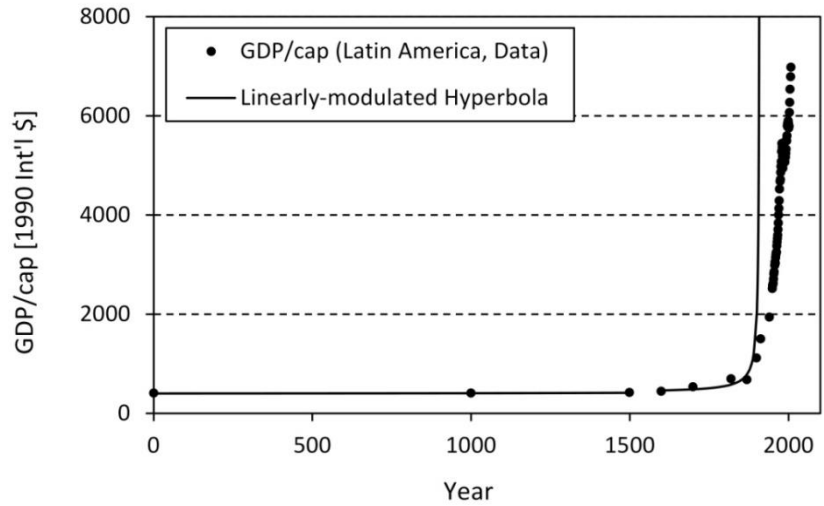

**Figure 16**. *Growth of income per capita in Latin America (Maddison, 2010; Nielsen, 2016g). In clear contradiction of Galor's claim supported by his distorted presentation of Maddison's data, growth of income per capita continued to increase along a nearly vertical trajectory close to the historical linearly-modulated hyperbolic distribution. The growth was not diverted to a distinctly different and gently-increasing trajectory as claimed by Galor (2005a, 2011; cf Figure 3).*

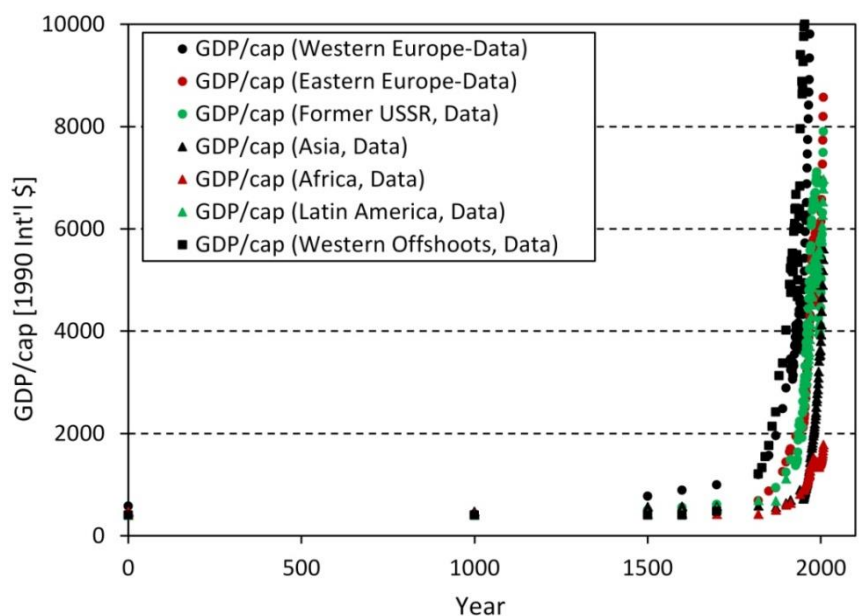

**Figure 17**. *Growth of income per capita in all regions, including Western offshoots (Maddison, 2010). Even without carrying out mathematical analysis of data it is clear that they all follow similar, nearly-vertical trajectories. The "mind-boggling" and "perplexing phenomenon of the Great Divergence" (Galor, 2005a, pp. 177, 220) has now been explained. The great divergence never happened. This mystery, as well as all his other "unresolved mysteries about the growth process" (Galor, 2005a, p. 220) has been created by the mind-boggling, perplexing and self-misleading manipulation of data.*

In Galor's distorted presentation of data shown in Figure 3, economic growth in various regions follows similar trajectories for a long time and then diverges to distinctly different trajectories. In the correct and undistorted presentation of data shown in Figure 17, economic growth in various regions follows similar trajectories all the time. Some regions are slower in their economic development but they all race in the same direction and along virtually the same trajectory. They do not fan out into distinctly different directions as claimed by Galor.

We do not have to explain the mechanism of the great divergence because the great divergence never happened. It is a feature created by the distorted presentation of data. If we want to explain the currently observed differences in the economic growth we should not be misguided by the Unified Growth Theory and we should not attempt to explain why different regions follow distinctly different trajectories, because they do not follow distinctly different trajectories. We should rather try to explain why different regions follow *similar* trajectories and why for some regions economic growth is faster while for other regions and it is slower.

## Geometric distortions

We shall now explain how Galor constructed his "unresolved mysteries about the growth process" (Galor, 2005a, p. 220): (1) his "mind-boggling" and "perplexing phenomenon of the Great Divergence" (Galor, 2005a, pp. 177, 220) and (2) his takeoffs from the alleged but non-existent stagnation to growth. For this purpose we can take any close family of distributions, which change slowly over a large range of independent variable and fast over its short range. We can use hyperbolic distributions, linearly-modulated hyperbolic distributions, a set of empirical distributions such as shown in Figures 10 and 17, or any other hyperbolic-like distributions. By their simple manipulation we can easily create Galor's "mysteries about the growth process" (Galor, 2005a, p. 220) but they will not be unresolved mysteries. They will not even be mysteries because we shall demonstrate and explain their origin. We shall demonstrate that these alleged mysteries do not represent unique properties of economic growth but the introduced by us disfigurations of hyperbolic-like distributions.

For our demonstration we have chosen three, closely-related linearly-modulated hyperbolic distributions shown in Figure 18. Like the historical income per capita distributions, each of these arbitrary distributions is represented by a ratio of two hyperbolic distributions. However, they have absolutely nothing to do with economic growth. They are purely mathematical functions $f(x)$, $g(x)$ and $h(x)$ where $x$ is an arbitrary independent variable. This variable could be time but it could be also anything else. The common feature of these distributions is that they start from approximately the same value at $x=0$, they increase monotonically (they are not characterised by sudden takeoffs at any time) and they increase to infinity within a small range of $x$ values. They do not diverge.

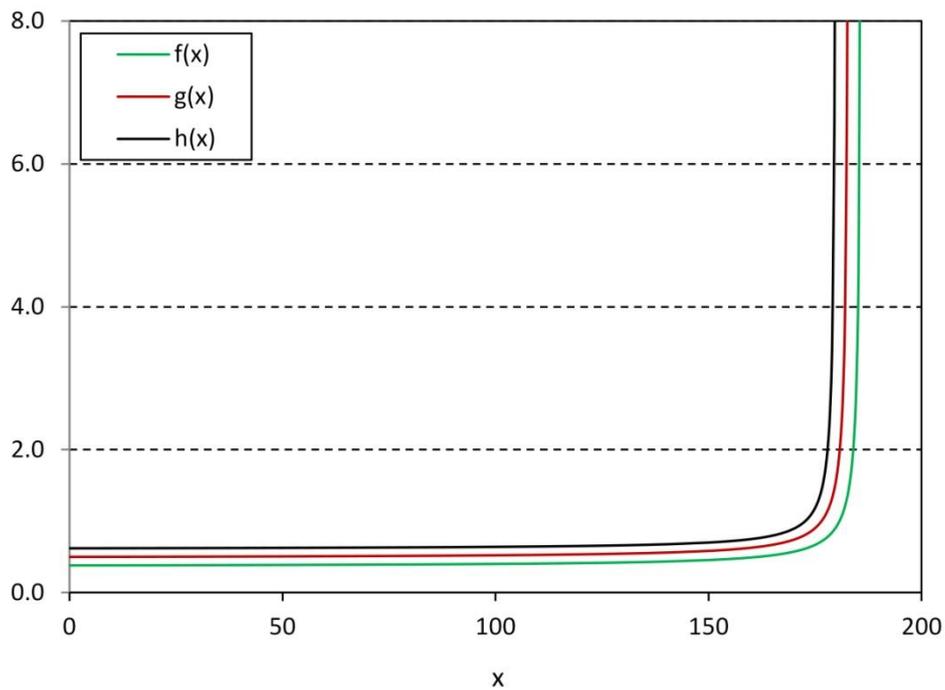

**Figure 18.** *Three arbitrarily-chosen, linearly-modulated, hyperbolic distributions, $f(x)$, $g(x)$ and $h(x)$. They increase* monotonically *from approximately the same value at $x=0$ to infinity within approximately the same time.*

However, if we follow Galor's example we can use these non-divergent and monotonically-increasing distributions and *construct a new set of distributions*, which will be diverging and which will be characterised by clear takeoffs. All we have to do is to select a few strategically-located points at certain constant *x*-values and join them by straight lines. This is precisely what Galor was doing repeatedly during the formulation of his Unified Growth Theory (Galor, 2005a, 2011) and in his other publications (Galor, 2005b, 2007, 2008a, 2008b, 2008c, 2010, 2012a, 2012b, 2012c; Galor and Moav, 2002; Snowdon & Galor, 2008).

We have selected three values of independent variable, $x=0$, $x=150$ and $x=179.6$, and by following Galor's example, we have connected the corresponding values of $f(x)$, $g(x)$ and $h(x)$ by straight lines. We have now constructed typical distributions used by Galor to formulate his Unified Growth Theory. We have also constructed the great divergence and the takeoffs. Results are shown in Figure 19.

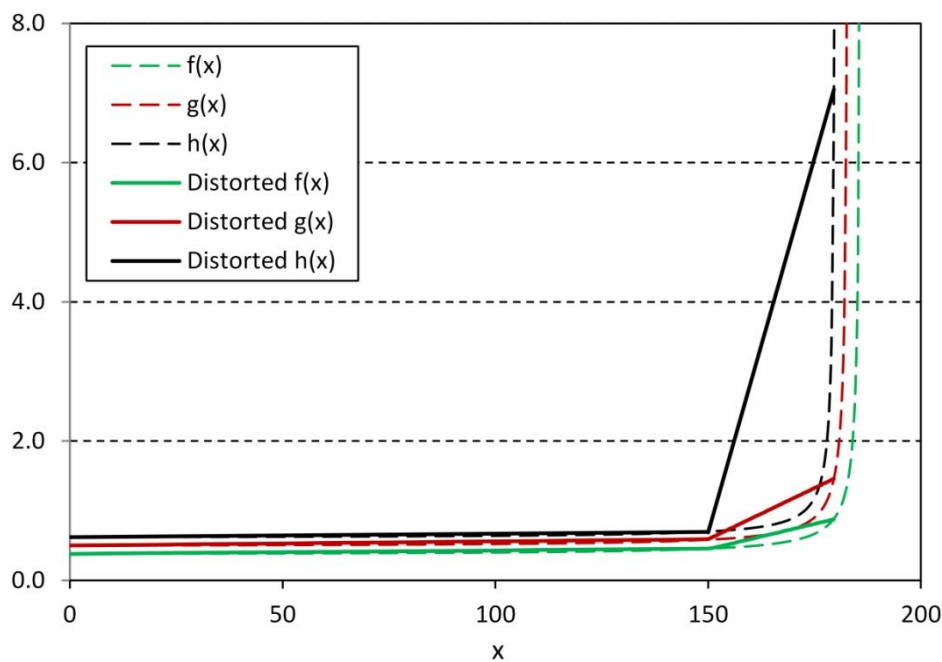

**Figure 19.** *This figure explains how the "mind-boggling" and "perplexing" phenomenon of the Great Divergence" (Galor, 2005a, pp. 177, 220) was invented by Galor and how he created his otherwise non-existent takeoffs from the non-existent stagnation to growth. By following his approach to research, the corresponding values of the purely-mathematical functions $f(x)$, $g(x)$ and $h(x)$ at $x=0$, $x=150$ and $x=179.6$ were joined by straight lines. The monotonically-increasing distributions are now replaced by distorted diagrams in much the same way as Maddison's data were replaced by Galor by his distorted diagrams. We have constructed the meaningless "mind-boggling" and "perplexing phenomenon of the Great Divergence" (Galor, 2005a, pp. 177, 220) preceded by the equally meaningless takeoffs at $x=150$.*

It would be incorrect to claim that our constructed distributions shown in Figure 19 represent the original distributions but Galor repeatedly and incorrectly uses his distorted diagrams as representing Maddison's data. His repeatedly used diagrams are the misrepresentations of data and his conclusions based on such diagrams or on quoting some isolated numbers selected from hyperbolic distributions are scientifically unacceptable and strongly misleading.

By using the constructed great divergence and the takeoffs shown in Figure 19 and by constructing more of such diagrams we could now create a unified growth theory describing properties of the distorted diagrams and insist that they represent mathematical properties of $f(x)$, $g(x)$ and $h(x)$ functions or the properties of other similar distributions. However, it would be naïve for us to expect that people familiar with mathematics would be impressed by our scholarly performance and by the mysteries we have created. It would be naïve to expect that they would accept our explanations of the claimed mathematical properties of hyperbolic-like distributions, and yet Galor expects that economists will accept his distorted representations of Maddison's data and his explanations of economic growth based on such repeatedly distorted presentations of data reinforced by the numerous quotations of well-selected and isolated numbers, which are supposed to represent a reliable empirical confirmation of his theory.

Like Galor, we could claim the existence of takeoffs from stagnation to growth for our mathematical, monotonically-increasing functions $f(x)$, $g(x)$ and $h(x)$. We could try to explain these phantom takeoffs by some fanciful mechanisms, but such explanations would be unacceptable because the original functions increase monotonically. They are not characterised by sudden takeoffs. These takeoffs do not exist. We have created them by distorting the original functions.

Like Galor, we could also claim the existence of the great divergence and try to explain it by some complicated mechanisms but again our claim and our explanations would be unacceptable because the original functions do not diverge. We have created the great divergence, which does not characterise the original functions but only their distorted representations. Like Galor, we could claim the existence of the "unresolved mysteries" (Galor, 2005a, p. 220) about mathematical functions but the only audience we could hope to impress would be people who are not familiar with mathematics.

Conclusions based on the distorted representations of mathematical distributions $f(x)$, $g(x)$ and $h(x)$ can be obviously rejected. Likewise, conclusions based on Galor's distorted representations of Maddison's data can be and even should be rejected. Galor presents many curious and seemingly logical stories about economic growth but his stories are either repeatedly contradicted by data (Nielsen, 2014, 2015a, 2016a, 2016b, 2016d, 2016e, 2016f, 2016g, 2016h) or have no convincing confirmation in data. They have to be accepted largely by faith. Stories of fiction can be also attractive, logical and convincing but they will remain stories of fiction.

It would be incorrect to claim that the distorted diagrams presented in Figure 19 represent the mathematical distributions $f(x)$, $g(x)$ and $h(x)$. Likewise, it would be incorrect to claim that the distorted diagrams used repeatedly by Galor in his Unified Growth Theory and in his other publications represent Maddison's data.

It would be incorrect to claim that the distorted diagrams presented in Figure 19 describe the mathematical functions $f(x)$, $g(x)$ and $h(x)$. Likewise, it would be incorrect to claim that the distorted diagrams presented by Galor in his Unified Growth Theory and in his many other publications describe economic growth. They describe the world of fiction.

As it is not expected that explanations of the properties of mathematical functions $f(x)$, $g(x)$ and $h(x)$ based on their distorted representations shown in Figure 19 would be ever accepted by people familiar with mathematics, so also it can be hardly expected that explanations of economic growth based on such distorted presentations of data as used by Galor in his Unified Growth Theory and in his other publications can be accepted by the scientific community.

## Discussion and conclusions

We have analysed Maddison's data (Maddison, 2001, 2010) and we have demonstrated that the great divergence claimed by Galor (2005a, 2010) and shown in Figure 3 never happened. Various regions are now on different levels of development but their economic growth did not diverge into distinctly different trajectories as claimed by Galor (see Figure 3). Their income per capita increases along similar, approximately vertical trajectories.

The disagreement between Galor's claim and the data can be demonstrated using the early Maddison's data (Maddison, 2001), which Galor used in their habitually distorted presentations during the formulation of his Unified Growth Theory (Galor, 2005a, 2010). However, the disagreement between his claims and the data becomes even more pronounced if we display the latest data of Maddison (2010), which were available to Galor before the publication of his book (Galor, 2011).

The data do not even have to be analysed mathematically to show that they contradict Galor's claim of the existence of the great divergence but their mathematical analysis is helpful. Galor's claims expressed in his Unified Growth Theory and in his other similar publications are based on his failure to adhere to the fundamental and indispensable principles of scientific investigation, which require that data should be rigorously analysed, that conclusions should not be based on impressions and that data should not be manipulated to support preconceived ideas. His theory and his claims and interpretations are scientifically unacceptable.

"The mind-boggling phenomenon of the Great Divergence in income per capita across regions of the world in the past two centuries, that accompanied the take-off from an epoch of stagnation to a state of sustained economic growth, presents additional unresolved mysteries about the growth process" (Galor, 2005a, p. 220). It is interesting how a single sentence can contain so much misinformation.

His mysteries have now been solved: he has created them by the manipulation of data.

The great divergence never happened and neither did the takeoffs from Malthusian stagnation to growth (Nielsen, 2014, 2015a, 2016a, 2016b, 2016d, 2016e, 2016f, 2016g, 2016h). Within the range of analysable data there was no stagnation and no claimed transition from stagnation to growth. Features described by Galor as takeoffs are not takeoffs but the natural continuations of monotonically-increasing hyperbolic distributions describing the growth of the GDP or population, or the natural continuations of monotonically-increasing linearly-modulated distributions describing the growth of the GDP/cap. Hyperbolic distributions or linearly-modulated hyperbolic distributions are slow over a long time and fast over a short time but they do not change suddenly from slow to fast at any time. They increase monotonically all the time.

Hyperbolic growth excludes the interpretations revolving around the concept of Malthusian stagnation and around takeoffs from stagnation to growth described usually as the escape from the Malthusian trap. The evidence contradicting such interpretations is overwhelming. It is remarkable that so many independent studies are in such perfect agreement: Maddison's data (Maddison, 2001, 2010) and their analysis (Nielsen, 2014, 2015a, 2016a, 2016d, 2016e, 2016f, 2016g, 2016h); the estimates of the size of human population not only during the AD era but also during the BC era (e.g. Biraben, 1980; Clark,1968; Cook,1960; Durand, 1967, 1974, 1977; Gallant, 1990; Haub, 1995; Livi-Bacci, 1997; McEvedy & Jones, 1978; Taeuber & Taeuber, 1949; Thomlinson, 1975; Trager, 1994) and their analysis (e.g. Kremer, 1993; Nielsen, 2016b; Kapitza, 2006); the discovery made by von Foerster, Mora and Amiot (1960) and similar identifications of hyperbolic growth by Podlazov (2002), Shklovskii (1962, 2002) and von Hoerner (1975).

According to Galor, the "differential timing of the take-off from stagnation to growth across countries, and the corresponding variations in the timing of the demographic transition, led to a great divergence in income per capita as well as population growth" (Galor, 2005a, p. 218). This is a good example how fiction can be created even in science. Non-existent takeoffs have been constructed by distorted presentations of data. These non-existent takeoffs were then used to explain the non-existent differential takeoffs, and now the same phantom takeoffs are used to explain the origin of the non-existent great divergence constructed by the manipulation of data.

Galor wonders about "the underlying driving forces that triggered the recent transition between these regimes and the associated phenomenon of the Great Divergence in income per capita across countries" (Galor, 2005a, pp. 174, 219). While it is interesting to study reasons for differences in the *level* of economic growth of various regions and countries, there is no need not to wonder about the underlying driving forces of the great divergence because the great divergence never happened.

Galor claims that "In the course of the 'Great Divergence' the ratio of GDP per capita between the richest region and the poorest region has widened considerably from a modest 3 : 1 ratio in 1820, to a 5 : 1 ratio in 1870, a 9 : 1 ratio in 1913, a 15 : 1 ratio in 1950, and a 18 : 1 ratio in 2001." (Galor, 2005a, p. 174). All these ratios are probably correct but the conclusion is incorrect because there was no great divergence.

This is a good example of being guided by impressions and of using them to draw hasty conclusions. Data have to be rigorously analysed. Using isolated numbers, as done repeatedly by Galor, is likely to lead to incorrect interpretations particularly if such use of isolated numbers is combined with the repeatedly distorted presentation of data, such as shown in Figures 1 and 3. Taking shortcuts and using them to draw hasty conclusions based usually on preconceived ideas and on wished-for interpretations does not represent the scientific process of investigation. The ratios listed by Galor do not prove the existence of the great divergence. We have already demonstrated that the great divergence never happened. The listed ratios represent nothing more than hyperbolic growth and different levels of development *along virtually identical trajectories*.

Current economic growth in various regions and countries is at different *levels* of development. For countries characterised by high human development, income per capita can be as high as tens of thousands of dollars but for countries characterised by low human development it can be about one hundred times lower (Nielsen, 2006). However, Maddison's data show that economic growth in all regions, without exception, is developing along *similar trajectories*.

Galor's interpretation of economic growth is potentially dangerous because it creates a false sense of security. He shows that gradients of the current economic-growth trajectories are in general small and consequently the imminent economic crisis is unlikely (see Figure 3).

However, data convey totally different information. Economic growth in *all* regions is now increasing rapidly along virtually vertical trajectories (see Figures 4-17). They resemble the historical linearly-modulated trajectories, which increase to infinity at a fixed time. For such trajectories, economic crisis can be expected because the growth has to be supported by excessively large per annum increase in the GDP per capita. The created stress can be too high to be manageable over a long time. There is also a danger of reaching quickly natural limits to growth.

Warning signs can be already seen in Eastern Europe, in countries of former USSR and in Africa (Figures 12, 13, 15). Their growth of income per capita suffered reversals but after a certain time it managed to recover and follow again the nearly vertical trajectories. Certain degree of instability can be also observed in Latin America (Figure 16).

The preferred option would be to follow now gently-increasing trajectories but all regions, without exception, appear to be caught up in the general frenzy to increase rapidly their per capita economic growth. When they are temporarily left behind they soon resume their hazardous race. Current trajectories do not increase to infinity at a fixed time but they increase to infinity in a short time, which is hardly a consolation.

All these important warning signs are not even noticed in the Unified Growth Theory. Unified Growth Theory suggests a prosperous future after an ages-long epoch of a hypothetical stagnation but the data show that the future of economic growth is approaching rapidly levels of unsustainability. It has been shown that the world economic growth follows unsustainable trajectory (Nielsen, 2015b). However, the analysis of Maddison's data presented here suggests that this is a common danger shared by *all* regions. There is not a single region, whose economic growth diverged to a safer trajectory.

The two opposite interpretations of economic growth have also essential impact on research activities. In order to explain Galor's great divergence we would have to explain why there was a transition to distinctly different trajectories of economic growth. Such attempts would be a waste financial and human resources and a waste of time because the great divergence never happened. What we have to explain is why different regions follow virtually the same trajectories and why they follow such potentially-hazardous, fast-increasing trajectories. Why there is such a strong desire to increase the GDP *per capita* so quickly everywhere and how to control these dangerous tendencies.

Galor claims that the "transitions from a Malthusian epoch to a state of sustained economic growth and the emergence of the Great Divergence have shaped the current growth process in the world economy" (Galor, 2005a, p. 221). They did not because there was no "emergence of the Great Divergence." Galor describes phantom features he created by his manipulation of data. These phantom features could not have shaped the past growth and they do not shape the current growth because they did not and do not exist. Galor describes the world of fiction and events that never happened. He then uses these non-existing phenomena to weave his theory around them.

Transitions from the Malthusian epoch of stagnation to a state of sustained economic growth never happened because there was no stagnation. Economic growth was sustained in the past because it followed steadily-increasing hyperbolic trajectories. Takeoffs, which are supposed to represent the claimed transitions from stagnation to growth, never happened (Nielsen, 2014, 2015a, 2016a, 2016b, 2016c, 2016d, 2016e, 2016f, 2016g, 2016h).

Galor's claims are based on distorted presentations of data and generally on repeatedly violating the fundamental principles of scientific investigation. They are based on impressions rather than on the rigorous scientific analysis of empirical evidence.

Galor claims that the "unified growth theory sheds light on the perplexing phenomenon of the Great Divergence in income per capita across regions of the world in the past two centuries" (Galor, 2005a, p. 177). If it does, then his theory is a fiction because the perplexing phenomenon of the great divergence never happened.

Why did we devote so much time for the discussion of Galor's Unified Growth Theory (Nielsen, 2014, 2015a, 2016a, 2016b, 2016c, 2016d, 2016e, 2016f, 2016g, 2016h)? One of the obvious reasons is that the aim of any scientific investigation is to discover the truth. Science looks for correct interpretations but Galor's theory is so obviously incorrect that it attracted immediate attention.

However, there is also another important reason: Galor's Unified Growth Theory is not only incorrect but also dangerously incorrect because it diverts attention from the urgent need to monitor, control and regulate the current economic growth. It would be unwise to accept his theory and his explanations because his incorrect explanations of the historical economic growth are linked strongly with the current economic growth, which affects our future.

Galor claims that after a long epoch of stagnation we are now in the regime of sustained economic growth. His theory also strongly suggests that the current economic growth is not only sustained but also sustainable because in general it follows slowly-increasing trajectories (see Figure 3). The future appears to be safe and secure.

However, precisely the same data, which he used during the formulation of his theory, show that the opposite is true. It was in the past that the economic growth was safe and secure but now it follows strongly hazardous trajectories. Recent analysis of the world economic growth also indicates that its future is insecure (Nielsen, 2015b), which is hardly surprising because our current combined ecological footprint is already significantly higher than the ecological capacity (WWF, 2010).

Why did Galor manipulate data? Why did he repeatedly present distorted diagrams to support his preconceived ideas? Why did he quote isolated and well-chosen but otherwise meaningless numbers to support his arguments? Why did he create such an elaborate work of fiction?

If we reject the suggestion that he did it all on purpose, then a possible explanation is that he did not know how to analyse data. However, this explanation is unconvincing because he appears to be familiar with mathematics. Anyone familiar with mathematics can quickly see the characteristic features of hyperbolic distributions in Maddion's data and the analysis of hyperbolic distributions is trivially simple (Nielsen, 2014). Maybe he was not sure how scientific research should be conducted but equally surprising is why his publications escaped the scrutiny of the peer-review system.

The most plausible explanation is probably that he was blinded by prejudice. It is what psychologists describe as the cascade behaviour, information cascade, informational avalanche, illusion of truth, illusory truth, illusion of familiarity, running with the pack, following the crowd, herding behaviour, bandwagons and path depending choice (Anderson & Holt, 1997; Begg, Anas & Farinacci, 1992; Bikhchandani, Hirshleifer & Welch, 1992, 1998; De Vany & Lee, 2008; De Vany & Walls, 1999; Easley & Kleinberg, 2010; Grebe, Schmid & Stiehler, 2008; Ondrias, 1999; Parks & Tooth, 2006; Ramsey, Raafat, Chater & Frith, 2009; Walden & Browne, 2003).

In the demographic and economic research this phenomenon is demonstrated by the reluctance to accept the compelling contradicting evidence simply because many demographers or economists would not agree with the contradicting evidence. It is safer to follow the crowd and run with the pack. Tradition is stronger than science and only an outsider who has not been blinded by prejudice and who is not afraid of being rejected by the crowd might dare to show that the accepted doctrines are incorrect. He or she is then likely to be ridiculed and rejected but science is a self-correcting discipline so sooner or later such resistance to accept the overwhelming empirical evidence will have to be broken, but it would be better for science and scientists if the required change in the paradigm is accepted sooner rather than later.

We now have a large body of data (Biraben, 1980; Clark,1968; Cook,1960; Durand, 1967, 1974, 1977; Gallant, 1990; Haub, 1995; Livi-Bacci, 1997; Maddison, 2001, 2010; McEvedy & Jones, 1978; Taeuber & Taeuber, 1949; Thomlinson, 1975; Trager, 1994), which we can use to improve our understanding of the economic growth and of the growth of human population. The correct understanding of these two processes might have essential impact on our future.

Mathematical analysis of data (Nielsen, 2014, 2015a, 2016a, 2016b, 2016d, 2016e, 2016f, 2016g, 2016h) reveals many interesting features, which call for further investigation. The past economic growth and the growth of human population were hyperbolic. Within the range of analysable data, which for the growth of human population extends down to 10,000 BC, there was no Malthusian stagnation. Hyperbolic growth was slow but remarkably steady. There were no transitions from stagnation to growth because there was no stagnation. There was no escape from the Malthusian trap in the economic growth or in the growth of population because there was no trap. There were no takeoffs from stagnation to growth claimed by Galor (2005a, 2011). There was no differential timing of takeoffs, claimed also by Galor, because there were no takeoffs.

We have demonstrated (Nielsen, 2016h) that there was no "sudden spurt in growth rates of output per capita" (Galor, 2005a, p. 220). Contrary to the similar claim made by Galor, there was also no sudden spurt in the growth rate of human population in the past 12,000 years (Nielsen, 2016b, 2016d). The "unresolved mysteries about the growth process" listed by Galor (2005a, p. 220) have now been solved. They were all created by him through the manipulation of data.

Industrial Revolution had no impact on changing the trajectories of economic growth and of the growth of population. There was no population explosion. What is perceived as takeoffs or explosions are just the natural continuations of hyperbolic growth (Nielsen, 2014). There was also no "mind-boggling" and "perplexing phenomenon of the Great Divergence in income per capita across regions of the world in the past two centuries" (Galor, 2005a, pp. 177, 220).

Recently, economic growth and the growth of human population started to be diverted to slower trajectories but these new trajectories continue to increase close to the historical hyperbolic trajectories. Analysis of data shows that not only the Unified Growth Theory but also the Demographic Transition Theory, which is based on similar assumptions, is repeatedly contradicted by empirical evidence (Nielsen, 2016c).

All these features suggest further investigation to answer important question about economic growth and about the growth of human population. Why the economic growth and the growth of human population were hyperbolic. Why the hyperbolic growth was so remarkably stable over such a long time in the past. Why was it not affected by many random forces that were no doubt present? Why the economic growth and population growth trajectories were not affected by the Industrial Revolution. The only exception where there is a correlation between the Industrial Revolution and the economic growth and the growth of population is Africa, the poorest region. This boosting can be explained by the colonisation of Africa rather than by the beneficial effects of Industrial Revolution. What models should be used to explain the historical hyperbolic economic growth and the growth of human population? What are the common features that link these two processes? Why was the economic growth and the growth of human population diverted relatively recently to new, non-hyperbolic trajectories? Are these new trajectories likely to change again into the apparently preferred hyperbolic growth? How to prevent such an undesirable event? What should be done to make the growth of population and economic growth sustainable? Much work needs to be done but it would unwise and potentially dangerous to be guided by the Unified Growth Theory.